\documentclass[12pt,aps,prd,preprint,showkeys]{revtex4}
\pdfoutput=1 
\usepackage[utf8x]{inputenc}
\usepackage{amsmath}
\usepackage{amsfonts}
\usepackage{amssymb}
\usepackage{graphicx}

\begin{document}

\title{ Flow and vorticity with varying chemical potential in relativistic heavy ion collisions}

\author{Abhisek Saha and Soma Sanyal } 
\affiliation{School of Physics, University of Hyderabad, Gachibowli, Hyderabad, India 500046}


\begin{abstract}
 {We study the vorticity patterns in relativistic heavy ion collisions with respect to the collision energy. The collision energy is related to the 
chemical potential used in the thermal - statistical models that assume approximate chemical equilibrium after the relativistic collision.  
We use the multiphase transport model (AMPT)  to study the vorticity in the initial parton phase as well as the final hadronic phase of the 
relativistic heavy ion collision. We find that as the chemical potential increases,the vortices are larger in size. Using different definitions of vorticity, we find that vorticity plays a greater role at lower collision energies than at higher collision energies. We also look at other effects of the flow patterns related to the bulk viscosity and the shear viscosity at different collision energies. We find that the shear viscosity obtained is almost a constant  with a small decrease at higher collision energies. We also look at the elliptic flow as it is related to viscous effects in the final stages after the collision. Our results indicate that the  viscosity plays a greater role at higher chemical potential and lower collision energies.}

\end{abstract}


\maketitle
\flushbottom

\section{Introduction}

The study of quark gluon plasma at various temperatures and baryon densities has shown remarkable progress in the past decade \cite{reviews,bass}. The quark 
gluon plasma has been observed experimentally in 
relativistic heavy ion collisions \cite{obser}. Experimental results show that depending upon various factors the quark gluon plasma can behave as an ideal fluid \cite{perfluid,zajc,huovinen}. While the context of an ideal fluid has been applied successfully for Au + Au collisions at high
energies, there is evidence that dissipative effects become more and more prevalent for lower collision energies \cite{heinz1}. The presence of collective flow dynamics has given rise to various theoretical and computational models. Hydrodynamic simulations have been developed to study transport coefficients 
close to the phase transition. It is still not clear though whether full thermal or chemical equilibrium can be achieved at all, in relativistic heavy ion collision. This has given rise to hybrid transport models. These combine many body dynamics with partonic and hadronic interactions as well important aspects of the phase transition dynamics.

Theoretically, the 
phase transition dynamics in the expanding fireball is described by a thermal-statistical model that assumes approximate chemical equilibrium.
For a particular collision energy, these models are characterized by two important parameters, the temperature $(T)$ and the chemical potential $(\mu_B)$.
The baryon potential can be parameterized with respect to the collision energy for these models. Most of these models are based on the hadronic resonance gas 
model \cite{HRGM}. These models also discuss the shear stress and the bulk viscosity of the fluid after the phase transition. In the case of heavy ion collisions, the peripheral collisions result in a large angular momentum in the initial stages.  This has led to studies of vorticity in heavy ion collisions. 
Though neither shear stress nor bulk viscosity is related to the generation of vorticity in the plasma, the evolution of the vorticity depends on the
 interaction of the shear stress and the bulk viscosity.  According to the hadronic resonance gas (HRG) models, both the bulk viscosity as well as the shear viscosity depends on the chemical potential (or baryon density) of the plasma. Since both these quantities are  responsible for the evolution of vorticity patterns in a viscous fluid, we are interested in studying the effect of shear viscosity on the vorticity patterns seen in the quark gluon plasma. We therefore  do a systemic study of vorticity generated in the reaction (x-z) plane of heavy ion collisions at different collision energies. Different collision energies are related to the chemical potential and therefore the baryon density. Since there are various definitions of vorticity, we plot the 
average vorticity for different cases. In all the cases, the average vorticity goes down with increase in the collision energies. We then use the definition of shear viscosity from the HRG models and calculate the shear viscosity for different collision energies. As different collision energies mean different chemical potentials, it appears that the shear viscosity does not depend very strongly on the chemical potential, i.e the baryon density of the quark gluon plasma. However, there is some dependence. It is higher at low collision energy (higher chemical potential), while it is lower at higher collision energy (lower chemical potential). Since the decrease in vorticity is greater than the decrease in shear viscosity with respect to the collision energy, we conclude that bulk viscosity plays a greater role in the evolution of the vorticity pattern than the shear viscosity. This is also predicted by most of the HRG models. So the predictions from the AMPT seem to agree with the predictions from the HRG models, even though they are very different models.

For our study we use a publicly available simulation code, the AMPT code which is a  multiphase transport model to generate the initial distribution of the particles \cite{ampt}. We then use a grid based simulation to obtain the vorticity in the partonic phase as well as in the hadronic phase. We do this for various different collision energies keeping the rest of the parameters constant. This essentially means that only the baryon chemical potential is different for the different cases as the collision energy is related to the baryon chemical potential. Our choice of the reaction plane is based on the fact that the angular momentum and the vorticity along the y - direction is much greater than in the other directions. This has been shown in the literature before \cite{jiang}. We have considered the standard coordinate system where the beam direction as $\hat{z}$ axis, the impact parameter direction is $\hat{x}$ axis, and the out-of-plane direction is $\hat{y}$ axis. 

We calculate the weighted kinetic vorticity, both non-relativistic as well as relativistic with respect to the collision energy. We find that it goes down with an increase in the collision energy. We also get similar results for the thermal vorticity. For the specific shear viscosity we find that it remains almost constant for different collision energies. There is a small decrease in its value as we go to higher collision energies but this decrease is less than the decrease in the vorticity. Since the shear viscosity also affects the elliptic flow, we obtain the elliptic flow from our simulations. We see that the elliptic flow does not show any strong changes for lower collision energies. Hence we conclude that it is the bulk viscous pressure  which plays a dominant role in the evolution of the vorticity patterns at lower collision energies. We hope that this investigation would lead to further studies into the vorticity patterns at lower collision energies. This may lead to better understanding of the transport properties of the quark gluon plasma at finite chemical potentials.

 While vorticity helps us to study viscous properties of the fluid that is generated, it has also been speculated that they can give rise to various anomalous effects. The presence of a strong electromagnetic field in the 
background which can couple to parity or charge odd domains in the plasma can lead to  anomalous transport phenomenon. This is known as the chiral magnetic effect (CME) \cite{cme}. Similar to this effect, one can also have so-called chiral vortical effect (CVE)\cite{cve}, which is the vortical analog of CME and represents the generation of vector and axial currents along the vorticity. Recently, the measurements performed by the STAR Collaboration at
RHIC \cite{rhiccme} and the ALICE Collaboration at LHC \cite{alicecme}
showed features consistent with these different effects.  We are currently interested only in the viscous effects of the quark gluon plasma. We plan to look at the chiral effects at a later stage.

In section II  we discuss dissipative hydrodynamics and vorticity in the reaction plane. In section III,we discuss the hadron resonance gas models and the definition of shear viscosity, that we use for our calculations. The definition of the shear viscosity that we use comes from the HRG model but we also show that it is similar to definitions that have been used previously in hydrodynamic models.  In section IV, we describe our simulations and in Section V, we present the results and discuss the implications of our results. Finally we present our conclusions in section VI.

\section{Vorticity in the reaction plane}

In recent times due to the emergence of various experimental results at lower collision energies, dissipative hydrodynamics has received a great deal of attention \cite{disshydro}. It has been investigated using various simulation codes such as ECHO - QGP \cite{echoqgp}, VISHNU \cite{perfluid,vishnu} etc. In many cases, different kinds of vorticity have been defined and various aspects of dissipative dynamics have been studied. Here we study the vorticity patterns with respect to the baryon density of the fluid formed in the collisions. Since as we discuss later, the baryon chemical potential is related to the collision energy, a study with respect to the collision energy would mean a study with respect to the baryon chemical potential.

In classical hydrodynamics, vorticity is the curl of the velocity field $\bf{v}$. For an ideal fluid this net vorticity is a constant and moves with the fluid. The vorticity essentially captures the rotational motion of the fluid. For a viscous fluid, the rotational motion will also introduce viscous stress between the fluid layers. As there are no boundary conditions in the case of the heavy ion collisions, the local vorticity patterns will be formed due to the viscous stress in the layers of the rotating fluid. In different studies of vorticity, the fluid has been modelled by the moment of inertia tensor to account for the rotating mass. In other cases relativistic hydrodynamics have been used to define the vorticity. However in relativistic hydrodynamics  vorticities can be defined in several ways \cite{echoqgp}. We study the classical vorticity, the kinematical vorticity  and the thermal vorticity with appropriate weights in our simulations. 

We would like to calculate the vorticity generated by the particles in the heavy ion collision based on their momentum. We do not invoke the moment of inertia tensor. One of the reasons for using the momentum to calculate the vorticity is the high Reynolds number of the system. As is well known the fluid during a heavy ion collision has a very high Reynolds number. In such cases the viscous stresses are relatively weak over most of the fluid. The action of the vorticity is then localized to thin layers of the rotating fluid. The momentum in the particles will give us this local vorticity. Vorticity is a three dimensional object defined by $\omega_i = \epsilon_{i,j,k} \frac{\partial v_k}{\partial x_j}$, but in several previous simulations it has been shown that the angular momentum in the out of plane direction (or the y - direction) is far greater than the angular momentum in the other directions. We initially calculate the vorticity at a particular collision energy for all three dimensions. We find that the vorticity component in the $x$ and $z$ directions are much smaller than the component in the $y$ direction. So for all the other collision energies we only calculate the vorticity in the reaction plane.  In the reaction plane the classical vorticity is given by,  
\begin{equation}
\omega_y = \omega_{xz} = \frac{1}{2} (\partial_z v_x - \partial_x v_z)
\end{equation}
Here, $v_x, v_y, v_z $ are the components of the velocity in the three directions and the factor $1/2$ is included to take care of the symmetrization. We have done the calculations for a classical vorticity even though the velocities are relativistic since there will be less fluctuations in the case of classical velocities. Moreover, it was shown in ref.\cite{csernai}, that the general nature of the results remain the same for the classical and the relativistic cases. Hence we do both the classical as well as the relativistic case.

In the relativistic case, the vorticity is given by, 
\begin{equation}
\omega_{\mu \nu} = \frac{1}{2} (\partial_\nu u_\mu - \partial_\mu u_\nu)
\end{equation}
where, $\partial_\mu = (\partial_0,\partial_x,\partial_y,\partial_z,) $ and 
$u_\mu = \gamma(1, -v_x, -v_y, -v_z)$. 
In the reaction plane, we would therefore have, 
\begin{equation}
\omega_{x z} = \frac{1}{2} \gamma (\partial_z v_x - \partial_x v_z) + \frac{1}{2}(v_x \partial_z \gamma - v_z\partial_x \gamma) 
\end{equation}
Before we go into the details of the simulation and the results we would like to discuss in more detail about the relativistic vorticity. Due to the presence of the $\gamma$ factor the relativistic vorticity is always greater than the classical vorticity \cite{csernai}. Hence  it is difficult to compare the magnitudes of the classical and relativistic vorticities. One may be able to compare them if they are weighted by certain factors as has been done in ref.\cite{csernai}. In fact we found that it is necessary to put some weight function to obtain the final vorticities. So we use the energy as a weight factor. The average vorticity for both the classical as well as the relativistic velocities are, 
\begin{equation}
<\omega_{x z} >~ = ~\frac{\sum\epsilon_{ij}\omega_{x z}^{ij}}{\sum\epsilon_{ij}} 
\end{equation}
Apart from the kinetic vorticity, the thermal vorticity plays an important role in heavy ion collisions. It is directly related to the particle polarization in the plasma. The thermal vorticity is defined by, 
\begin{equation}
\omega_{\mu \nu}^T~ = ~\frac{1}{2} (\partial_\nu \beta_\mu - \partial_\mu \beta_\nu)
\end{equation}
Here $\beta_\mu = \frac{u_\mu}{T}$. Here $T$ is the local temperature.  Again, as in the previous cases, we obtain $<\omega_{\mu \nu}^T>$ using the energy as the weight factor. 

The local vorticity depends on the velocity field at a particular instant and will evolve as the velocity field evolves. In general fluid dynamics the bulk viscosity ($\Pi$) is ignored and the shear viscosity $(\eta)$ is considered to be a constant to simplify the Navier Stokes equation.  An important concept that has not been taken into consideration in all the previous studies of vorticity in relativistic heavy ion collision is the concept of the dissipation function $\phi_\mu$. This function is difficult to define in this context however it is proportional to the bulk viscosity. An increase in the bulk viscosity will cause an increase in the dissipation function. This means that there will be more dissipation in the velocity field. As the local vorticity will depend on the velocity field, at higher bulk viscosities the vorticity lines will also be far more spread out. The effect of dissipation will be reflected in the vorticity patterns at the different collision energies. Since the dissipation function itself cannot be plotted, we obtain the average vorticities and plot them with respect to the collision energy.         

\section{Coefficient of Shear viscosity}

Apart from the bulk viscosity, the shear viscosity is also important in viscous flows. We obtain the definition of the shear viscosity from the hadron resonance gas models. In these models, the thermodynamic potential is given by, 
\begin{equation}
log(Z,\beta,\mu_B)= \int dm (\rho_b(m) log Z_b(\beta,\mu_B)+ \\
(\rho_f (m) log Z_f (\beta,\mu_B))
\end{equation}      
Here, the gas of hadrons occupy a volume $V$, at a temperature $\beta^{-1}$ and chemical potential $\mu_B$ (representing the finite baryon density). $Z_b$ and $Z_f$ are the partition functions of bosons and fermions respectively with mass $m$, while $\rho_b$ and $\rho_f$ are the densities of the bosons and fermions respectively. These models show a remarkable agreement with lattice QCD results for low temperatures. These models have been used to estimate coefficients of bulk viscosity for strongly interacting matter. Bulk viscosity coefficient is an important parameter as it increases near the 
 transition temperature. This has been observed in several models including the chiral perturbation theory \cite{cpt}, quasi particle models\cite{qpm} as well as Nambu–Jona-Lasinio model\cite{njl}. The HRG models also define the shear viscosity. In a recent work, Kadam and Mishra generalized the viscosity co-effecients of the hadron resonance gas model to include finite chemical potential effects \cite{mishra}. We use their definition of shear viscosity in our simulations. The shear viscosity is given by, 
\begin{equation}
\eta = \frac{5}{64 \sqrt{8}r^2} \Sigma_i <|p|> \frac{n_i}{n}
\label{shear}
\end{equation}
 Here $n_i$ is the number density of the $i$ th particle while $r$ is the radius of the particles concerned. The AMPT simulations allow us to identify the different particles. Instead of taking all the particles, we find the shear viscosity for the neutrons and the protons separately. We also find the shear viscosity of some other particles such as pions and $\Lambda$ hyperons separately. The general nature of the change in shear viscosity is the same for the different particles. It is only the magnitude which is different as the number of particles vary considerably.  The coefficient of shear viscosity is also the most studied transport coefficient in relativistic heavy ion collisions. Since it is primarily responsible for the equilibration of the momentum anisotropy, it can be estimated from the elliptic flow velocity. As we are interested in vorticity patterns we also study the elliptic flow velocity at different collision energies. The elliptic flow velocity will reflect the change in the shear viscosity more than the bulk viscosity. This will help us to understand which of the two viscosities play a greater role in the vorticity seen in the quark gluon fluid.

We obtain the coefficient of shear viscosity at different collision energies. 
The different collision energies can be related to the chemical potential.  
Based on the fact that all results on particle multiplicities are consistent with the assumption of chemical equilibrium in the final-state of the fireball that is produced, it is possible to parameterize the energy dependence of the baryon chemical potential by the relation \cite{Cleymans},
\begin{equation}
\mu_B(\sqrt{s}) = \frac{d}{(1+e\sqrt s)}
\label{eqn:roots}
\end{equation} 
with $d = 1.308 \pm 0.028 GeV$ and $e = 0.273 \pm 0.008 GeV^{-1}$. 
This parameterization is based on results obtained from different groups over a wide range of energies and therefore should be used with caution.
However, it is an established parameterization of freeze-out conditions at different energies. Since we are not fitting any data using this equation hence we feel that the equation is sufficient for our purposes. 
Our motivation is to understand the role of viscous stress in generating the vorticity patterns in the partonic fluid. This will mostly involve a qualitative description of the relation between the collision energy and the baryon chemical potential rather than a quantitative one.      

There have been other definitions of shear viscosities which have been used to show the dependence of viscosity on the baryon chemical potential ($\mu$). However, in  these cases, $\mu$ in an input parameter \cite{viscosityAMPT}. It is difficult to calculate the chemical potential $\mu$, unless we use eqn.\ref{eqn:roots}. Since as mentioned before the relation between the collision energy and the baryon chemical potential is not so rigorous, we prefer to calculate the viscosity coefficient from the particle velocities. The particle velocities for the individual particles are easily available as an output of the AMPT model.

\section{Simulations}

For our initial condition, we use the publicly available multiphase transport model, also known as the AMPT model \cite{ampt}.This is a hybrid 
type of transport model. The initial conditions of AMPT are obtained from the heavy-ion jet interaction generator (HIJING) \cite{hijing} and the 
scattering properties of the partons are fixed by Zhang’s parton cascade (ZPC) model \cite{zpc}. This model has been extensively used to 
investigate various transport properties of the heavy ion collisions \cite{sun}. The reason for choosing this model is that it incorporates both the partonic part as well as the hadronic part.  We are interested in studying the vorticity of the  particle 
flow both in the partonic as well as the hadronic part. The AMPT model has a string melting version as well as the default version. Though we have studied both the versions, results have been presented for the string melting version only. The parameters that we use in our simulations have been used previously in the AMPT model to study vorticity in the $(x-\eta)$ plane \cite{jiang}. The parameters have been used for a study that has reasonably reproduced the yields, transverse momentum spectra, and elliptic flow for low-$p_T$ pions and kaons in central and midcentral $Au + Au$ collisions at collision energies of 200 GeV \cite{lin}.

We first concentrate on the partonic stage. In the AMPT model, the impact parameter axis is the
x axis and the beam axis is the z axis, with the incoming
nucleus centered at $x = b/2 > 0 (x = −b/2 < 0)$ for positive
(negative) longitudinal momentum. This means that the initial total angular momentum is primarily along the y - direction. As mentioned before, that is the reason why we are only concentrating on the component of the angular momentum in the reaction plane. The output of the AMPT gives the particles space-time coordinates and three momentum at freeze-out. An event in the AMPT simulation generates thousands of
particles, however this amount is not large enough to generate a
smooth momentum distribution. Since we are interested in calculating the vorticity, a smooth momentum distribution is important for us. So we have 
to generate a very large number of events with the same parameters and average over them. 

Another crucial aspect of our simulation is the grid or cell size. We need to choose the cell size such that each cell has significant number of particles in them. We set the impact parameter at a fixed value of $b = 7 fm $ and choose a cell size of 0.5 fm in each direction. For each cell, 
the average momentum and the energy is calculated and then the velocity is extracted from these two values by taking $\frac{\langle \overrightarrow{p} \rangle}{\langle \epsilon \rangle}$. 
A similar method was followed in reference \cite{jiang}. However, unlike the analysis in Jiang et. al., we work in the reaction plane throughout 
this paper. We also do not dwell on the vorticity patterns at a given fixed energy since they have been analyzed previously. Jiang et. al have analyzed 
the pattern in the $(x-\eta)$ plane and hence their patterns are different from ours. Deng et. al \cite{deng} have studied it in the (x-y) plane. Though 
they have looked at two different collision energies both of these energies are very high. We look at the vorticity patterns at lower collision energies. 
Csernai et. al have looked at the vorticity in the (x-z) plane, using CFD simulations. Overall our vorticity patterns match with them, but due to 
the input that we obtain from the AMPT program, our vorticity patterns show more details.

\section{Results and Discussion}

\subsection{The initial partonic stage}

We have obtained the vorticity patterns in the reaction (x-z) plane at an collision energy between $200$ GeV and $20$ GeV. Here we do not present all the 
energies but only give selected patterns of vorticity to show  the difference in the pattern in the low energy collisions. Fig 1 and Fig 2 show 
the vorticity patterns at $200$ GeV and $100$ GeV. The  vorticity in both the plots range between -0.06 to 0.08. However, we see that the vortex lines 
indicate distinct contours around the vortices formed. Fig 3 gives the vorticity pattern at $20$ GeV. Here the contours are far more spaced out than in the previous two figures.  
\begin{figure}
\includegraphics[width = 86mm]{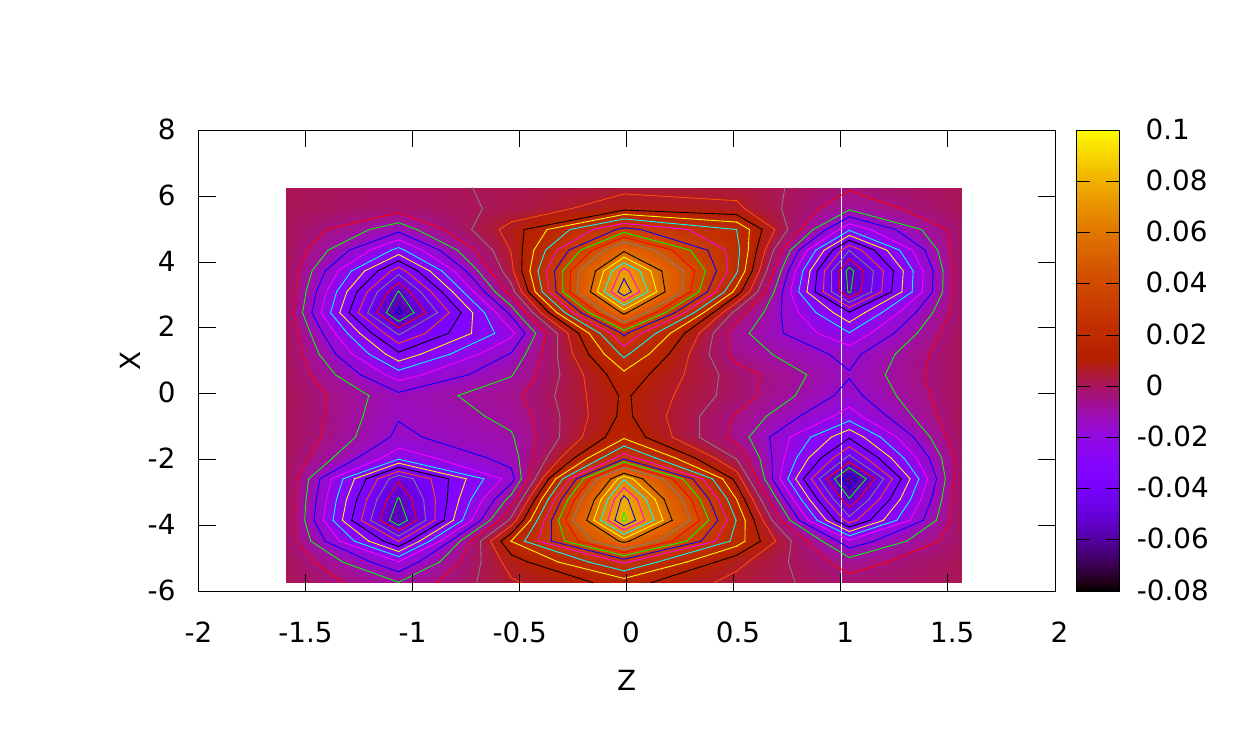}
\caption{Vorticity distribution in the reaction (x-z) plane at collision energy of 200 GeV for partons according to the non-relativistic definition }
\end{figure}
\begin{figure}
\includegraphics[width = 86mm]{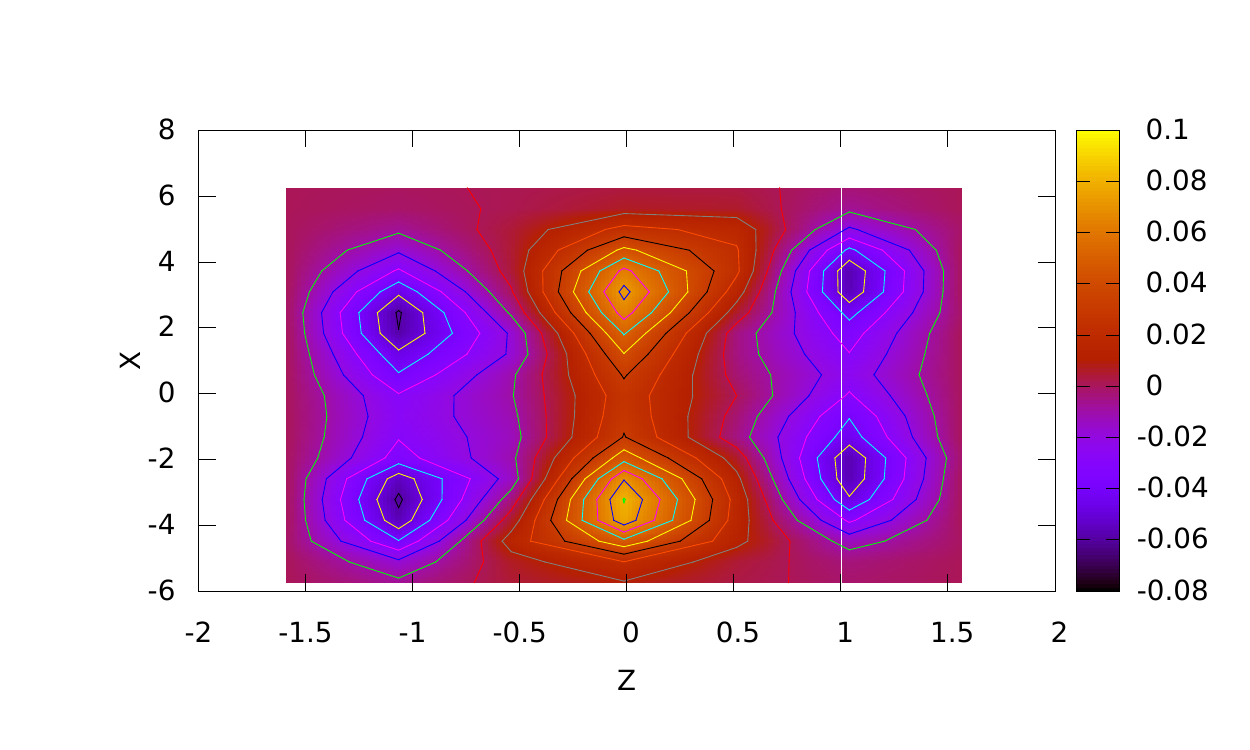}
\caption{Vorticity distribution in the reaction (x-z) plane at  collision energy of 100 GeV for partons according to the non-relativistic definition  }
\end{figure}

\begin{figure}
\includegraphics[width = 86mm]{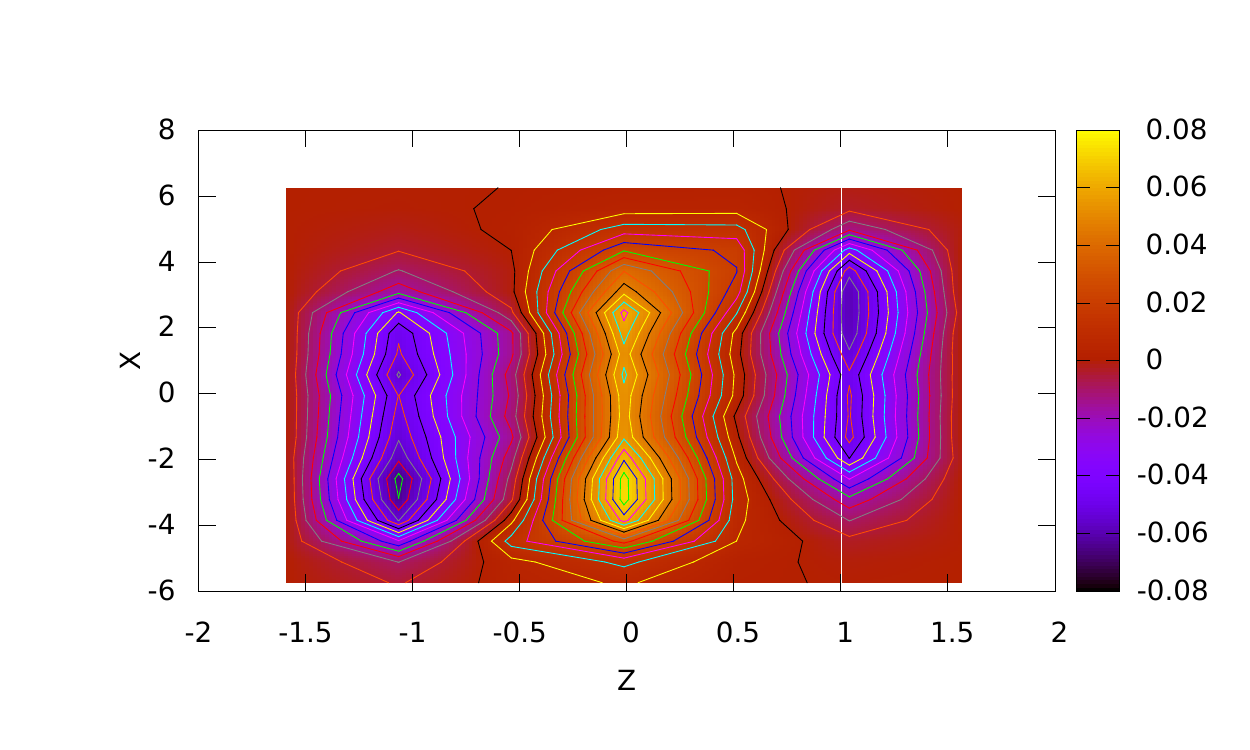}
\caption{ Vorticity distribution in the reaction (x-z) plane at  collision energy of 20 GeV for partons according to the non-relativistic definition}
\end{figure}

As the collision energy decreases, the chemical potential $\mu_B$ increases as per equation \ref{eqn:roots}. The vorticity formed tends to be circular if the angular momentum is higher and strain due to the bulk viscous pressure is lower. As the strain due to the bulk viscous pressure around the fluid increases, the vortices tend to spread out and become more elliptical in nature. The spread out elliptical vortices are due to the dissipation function which depends on the bulk viscosity. This is reflected in the simulations that we have performed.

We next look at the same energies for the relativistic vorticity.As expected, the values of the vorticity are much higher due to the relativistic $\gamma$ factor. As the particles move with velocities close to that of light, the relativistic  $\gamma$ factor is often of the order of $10^2$.
\begin{figure}
\includegraphics[width = 86mm]{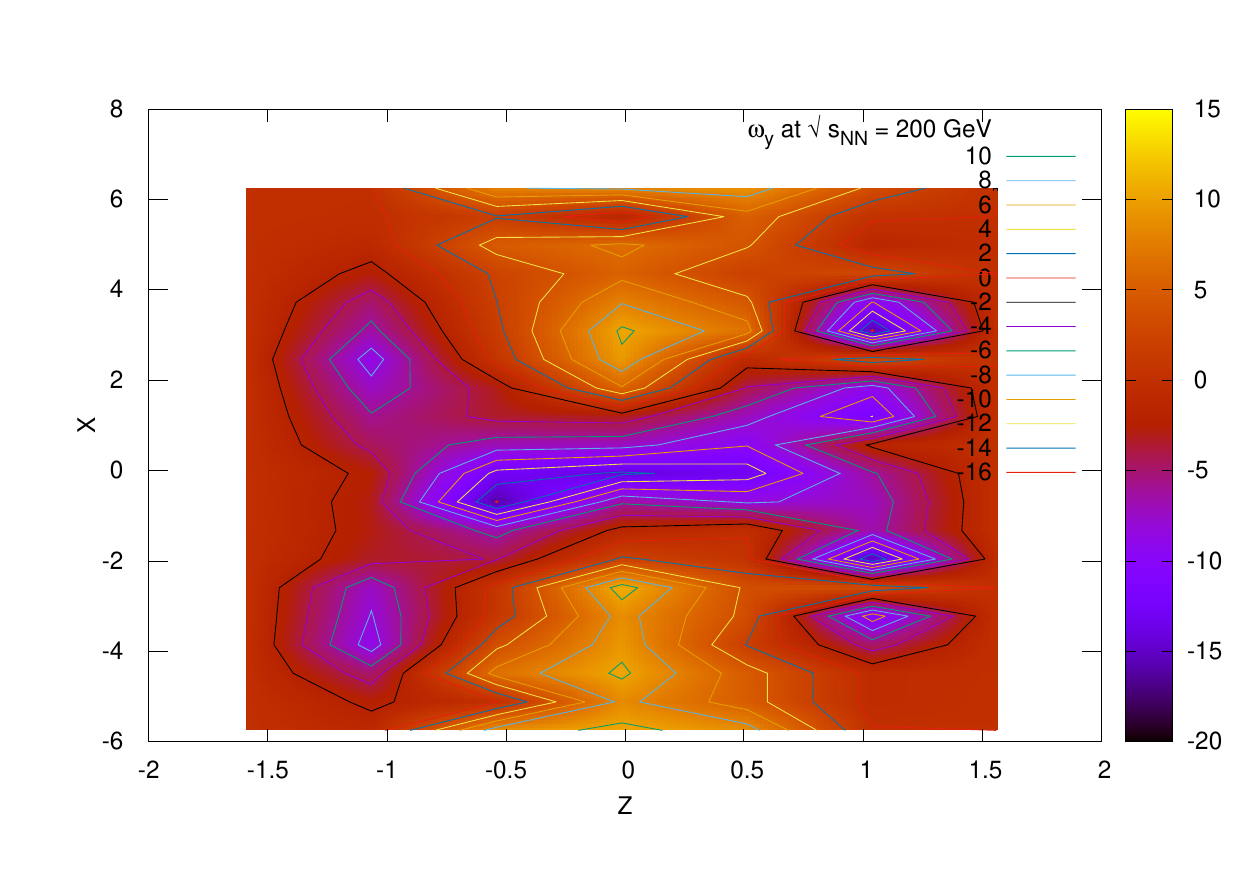}
\caption{ Kinetic vorticity distribution in the reaction (x-z) plane at an collision energy of 200 GeV for relativistic partons. }
\end{figure}

\begin{figure}
\includegraphics[width = 86mm]{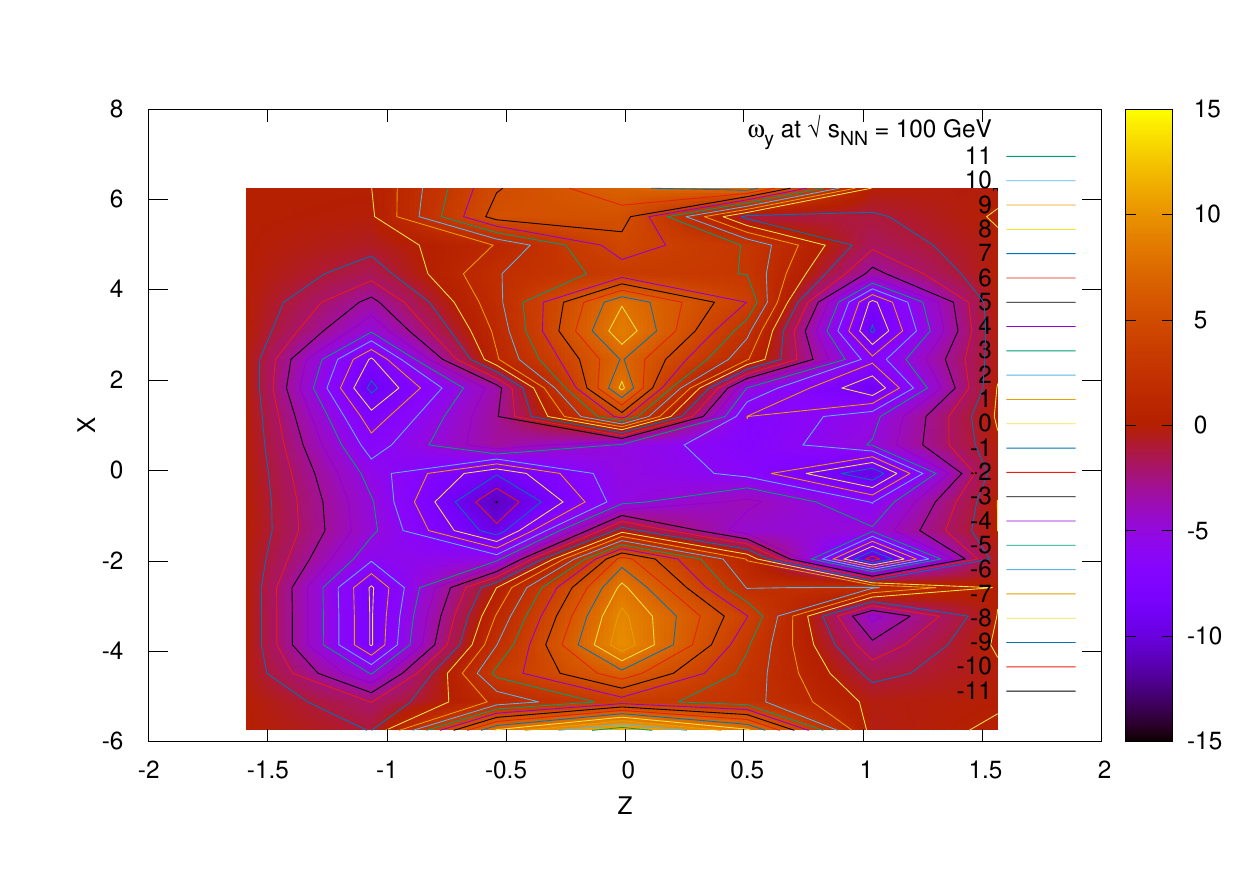}
\caption{ Kinetic vorticity distribution in the reaction (x-z) plane at an collision energy of 100 GeV for relativistic partons. }
\end{figure}

\begin{figure}
\includegraphics[width = 86mm]{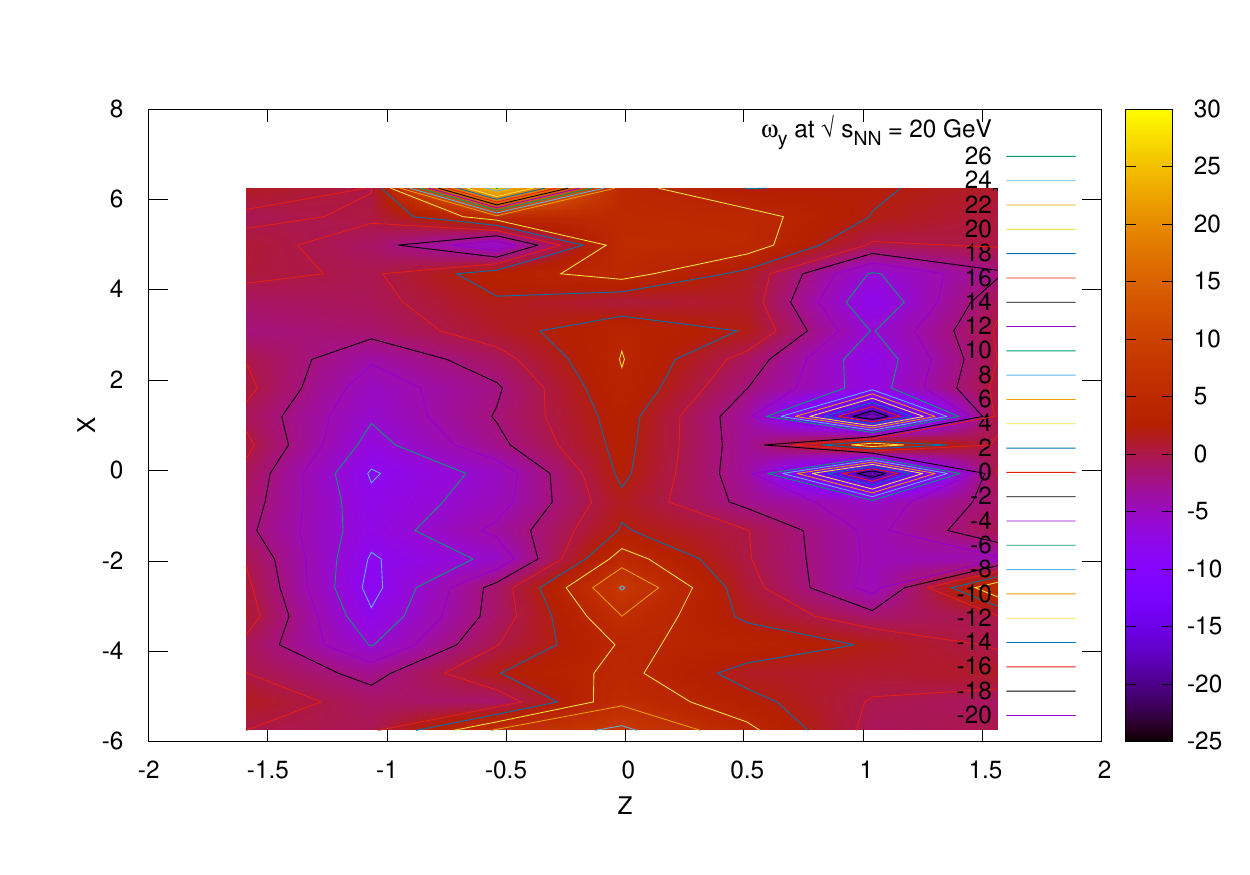}
\caption{ Kinetic vorticity distribution in the reaction (x-z) plane at an collision energy of 20 GeV for relativistic partons. }
\end{figure}

Though there is a general similarity between the vorticity patterns, there are also several dissimilarities. The most important difference is noticed at the 
center of the pattern. Also to be noticed, is that the differences are stronger at lower collision energies. Now we present some patterns for the thermal vorticity. 
Thermal vorticity is also required for studying the polarization of the particles, so we give the thermal vorticity patterns for the same energies. 

\begin{figure}
\includegraphics[width = 86mm]{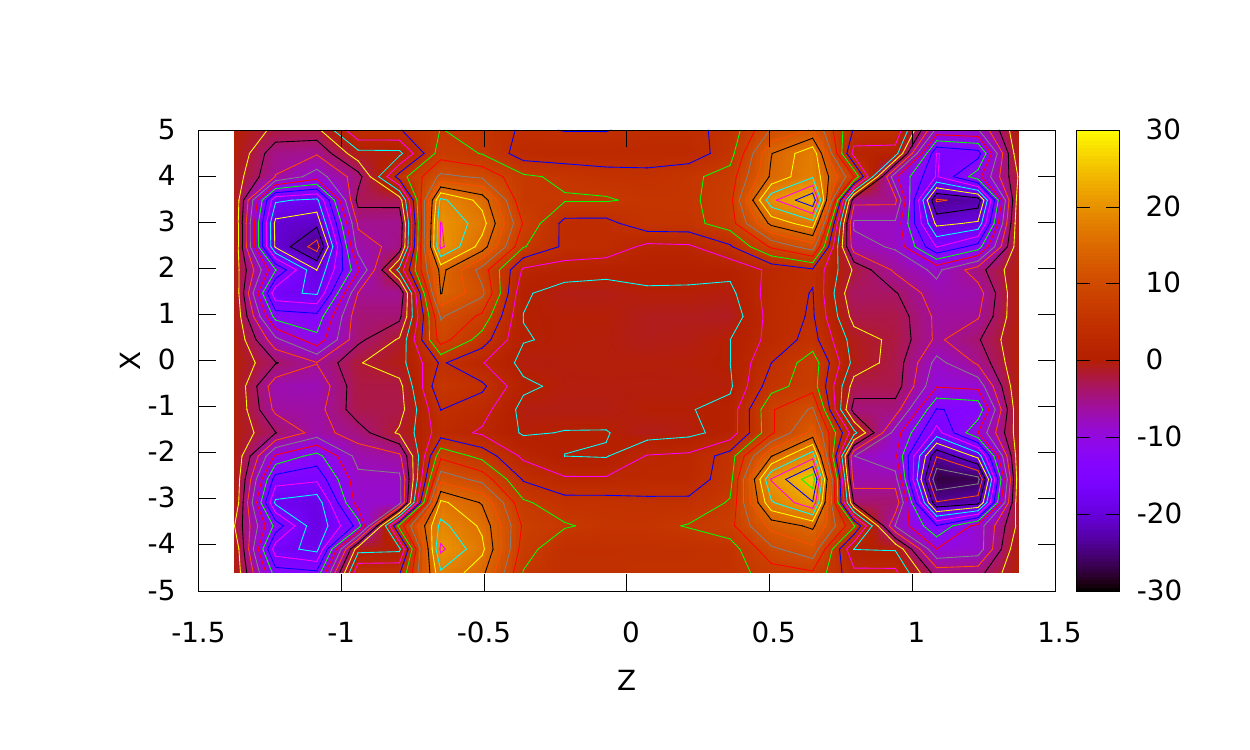}
\caption{ Thermal vorticity distribution in the reaction (x-z) plane at an collision energy of 200 GeV for partons. }
\end{figure}

\begin{figure}
\includegraphics[width = 86mm]{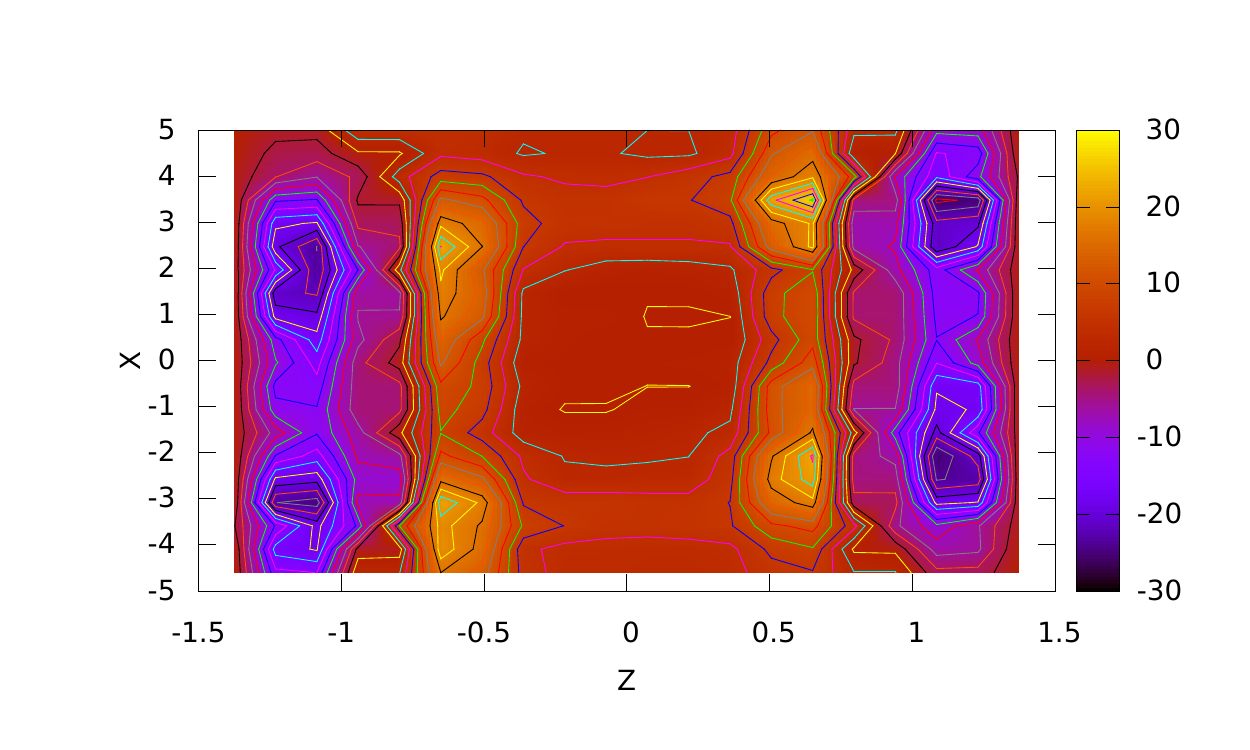}
\caption{ Thermal vorticity distribution in the reaction (x-z) plane at an collision energy of 100 GeV for partons. }
\end{figure}

\begin{figure}
\includegraphics[width = 86mm]{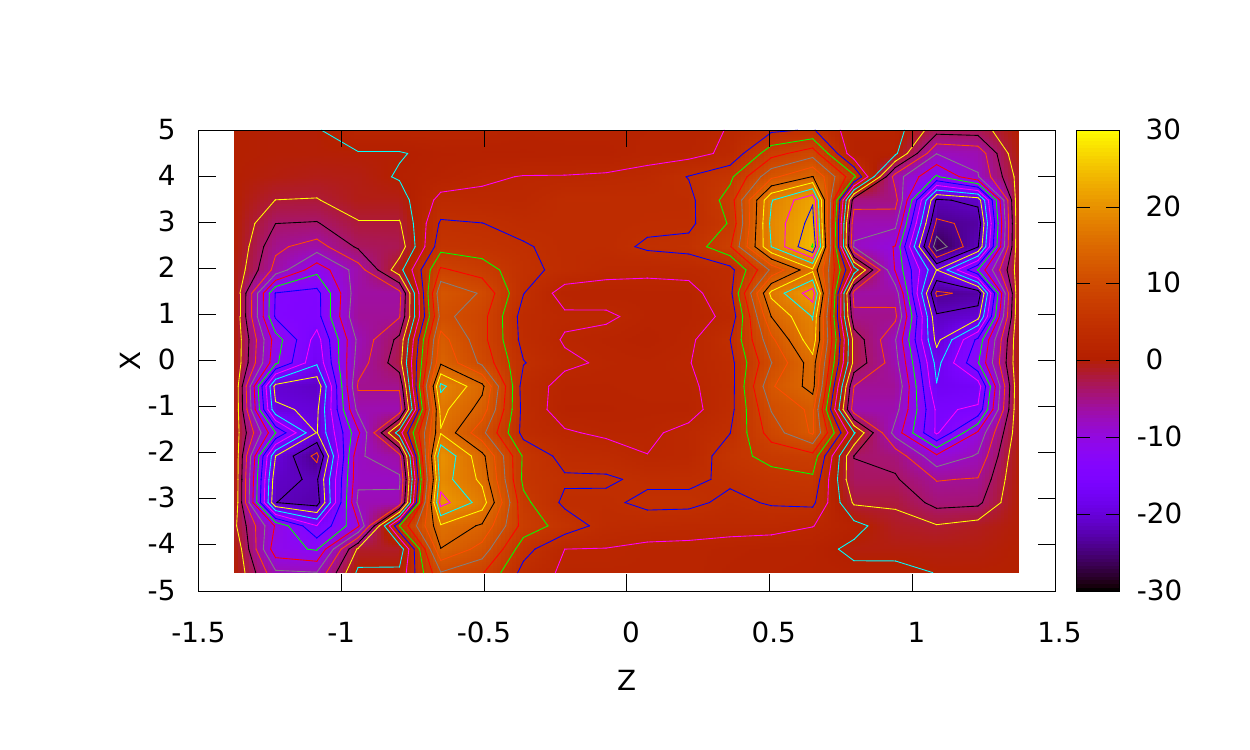}
\caption{ Thermal vorticity distribution in the reaction (x-z) plane at an collision energy of 20 GeV for partons. }
\end{figure}

In all the cases,  we observe that the vorticity spreads out as we go to lower and lower collision energies. So it can be concluded that as we go down in collision energies the vorticity pattern is more and more diffuse.  

There are some differences  between the non - relativistic pattern and the relativistic pattern. We notice that in the non-relativistic case the central area has close to zero vorticity for the higher collision energies. However, for the relativistic case, the central region has strong fluctuations and has a finite vorticity. The positive and negative numbers refer to opposite directions of the $\omega_{xz}$ vector. This happens for higher energy collisions. For the collision energy of $20$ GeV we see that the relativistic and the non-relativistic patterns appear to be similar but have very different magnitudes with the relativistic one having far more fluctuations than the non relativistic one.  
For the thermal vorticity, the patterns are similar to the non- relativistic case but there are more fluctuations. However, the general trend that the vortices are spread out at lower collision energies is also seen in this case.

\subsection{The final hadronic stage}

In the hadronic stage the net vorticity is lower than the partonic stage. 
This is expected as the initial fireball consisting of partons has more angular momentum. In the string melting scenario of the AMPT that we have used, the hadronization occurs via a simple quark coalescence model after the partons stop interacting \cite{ampt}. The three momenta is conserved but the hadrons will have higher masses than the partons. This reduces the net vorticity in the hadrons. 

\begin{figure}
\includegraphics[width = 86mm]{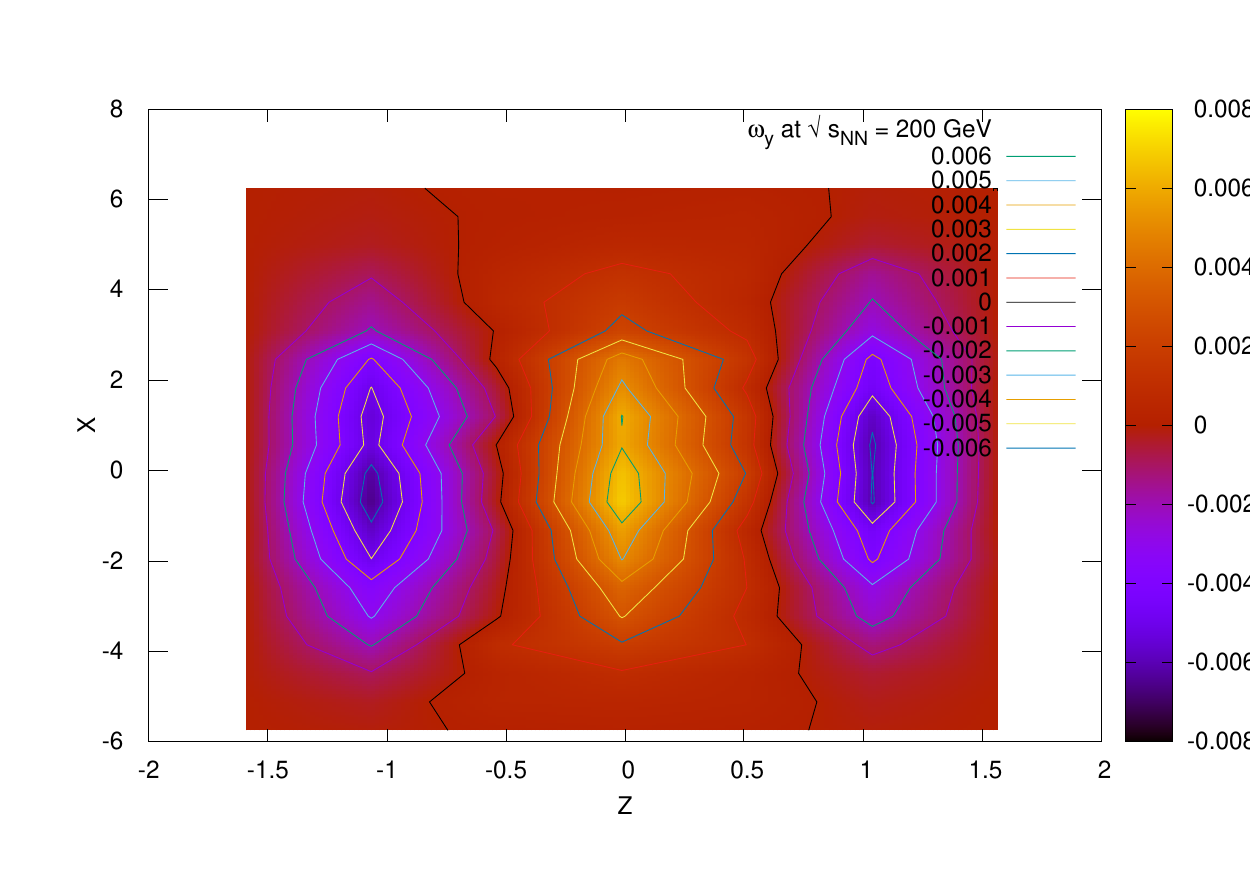}
\caption{ Vorticity distribution in the reaction (x-z) plane at an collision energy of 200 GeV for the hadronic phase }
\end{figure}

\begin{figure}
\includegraphics[width = 86mm]{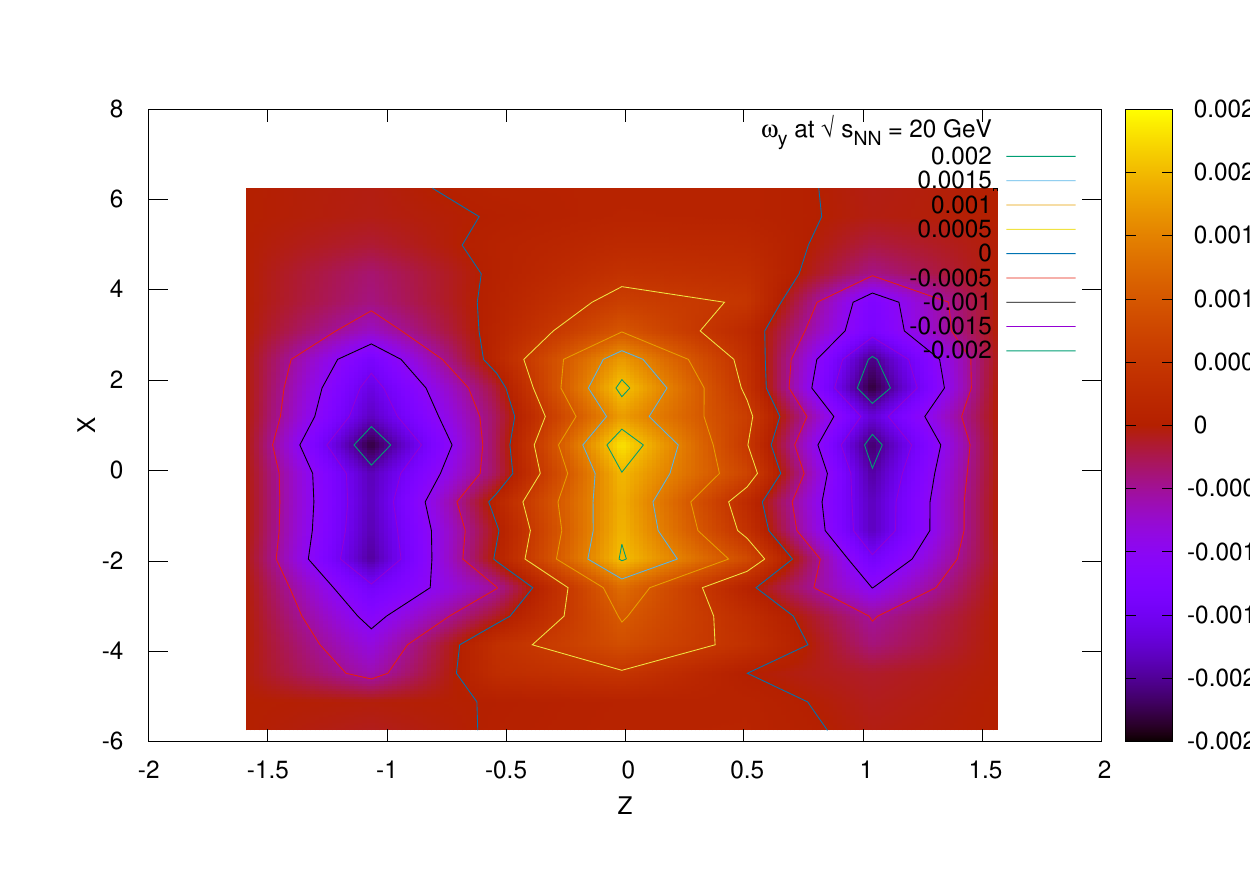}
\caption{Vorticity distribution in the reaction (x-z) plane at an collision energy of 20 GeV for the hadronic phase}
\end{figure}
The pattern is also spread out even at 200 GeV (fig 10). 
 Though there is a difference between spread in 200 GeV and the spread in 20 GeV (fig 11), it is not as distinct as in the partonic stage. However, the magnitude of the vorticity has become much smaller. In the relativistic case, the fluctuations are very large and no visible pattern is seen.

\subsection{Shear viscosity dependence on collision energy}
 
As mentioned previously, the vorticity patterns depend on the bulk viscous pressure and the shear viscosity. While the bulk viscous pressure increases with increasing baryon chemical potential, the specific shear viscosity  tends to  remain constant with increasing baryon chemical potential. We now calculate the coefficient of shear viscosity and plot it for various collision energies. We have defined shear viscosity in eqn.\ref{shear}. Though it is defined over all the particles, we do the calculations separately over various particles. As the right hand side is a summation, an average over all the particles would be of the same order of magnitude as the individual particles unless they differ from each other drastically. Since the number of neutrons, protons and pions
are greater in the output of the AMPT, we calculate the specific shear viscosity for these particles. The radius of the particles play an important role in obtaining the magnitude of the viscosity. We take the standard radius of the particles from the Particle Data book. 
Fig 12. shows the viscosity of the neutrons and the protons. We have checked the values by taking different values of $r$, all based on the limits  provided by the particle data group, we find that the general nature of the graph remains the same, as well as the general order of magnitude. 
   
\begin{figure}
\includegraphics[width = 86mm]{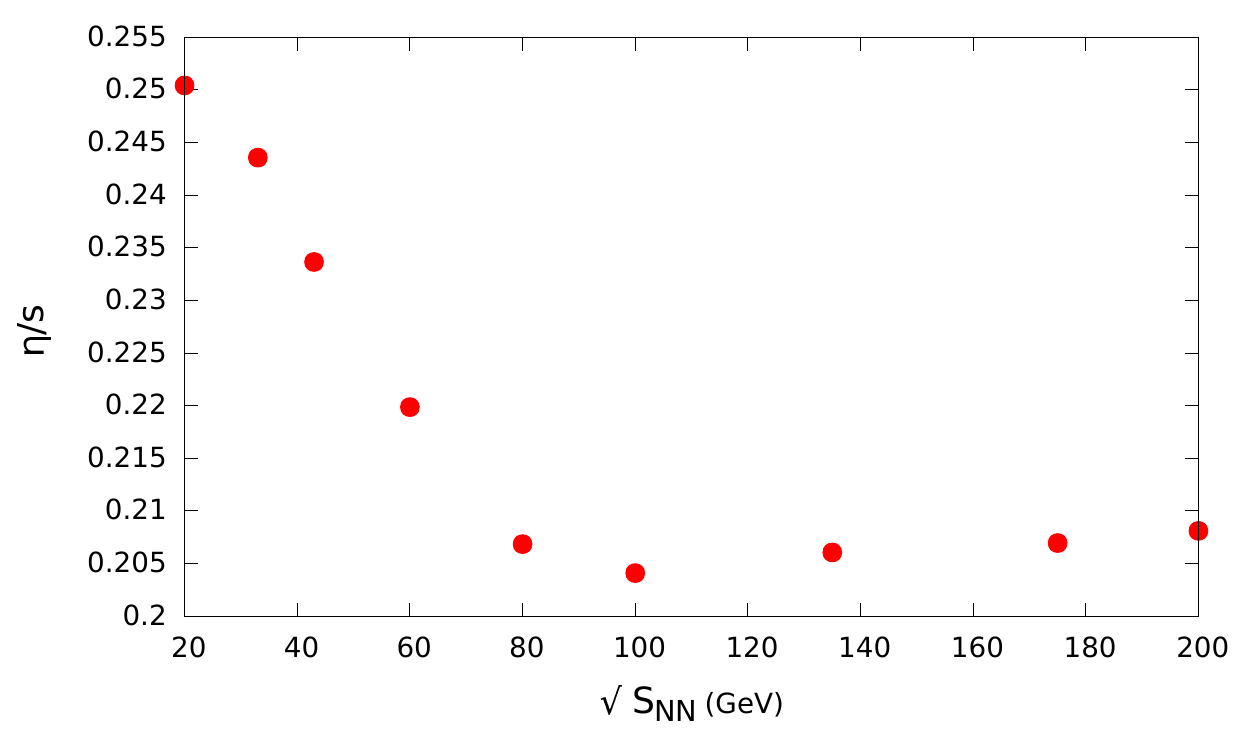}
\caption{ The  specific shear viscosity at different collision energies for neutrons, protons and their antiparticles. }
\end{figure}

As seen from the graph the specific shear viscosity is highest at lower collision energies which corresponds to higher baryon chemical potentials. However, it becomes nearly constant beyond $80$ GeV. 
This is further manifested in fig 13, where we plot the viscosity coefficient for pions and $\Lambda$  hyperons also. 
\begin{figure}
\includegraphics[width = 86mm]{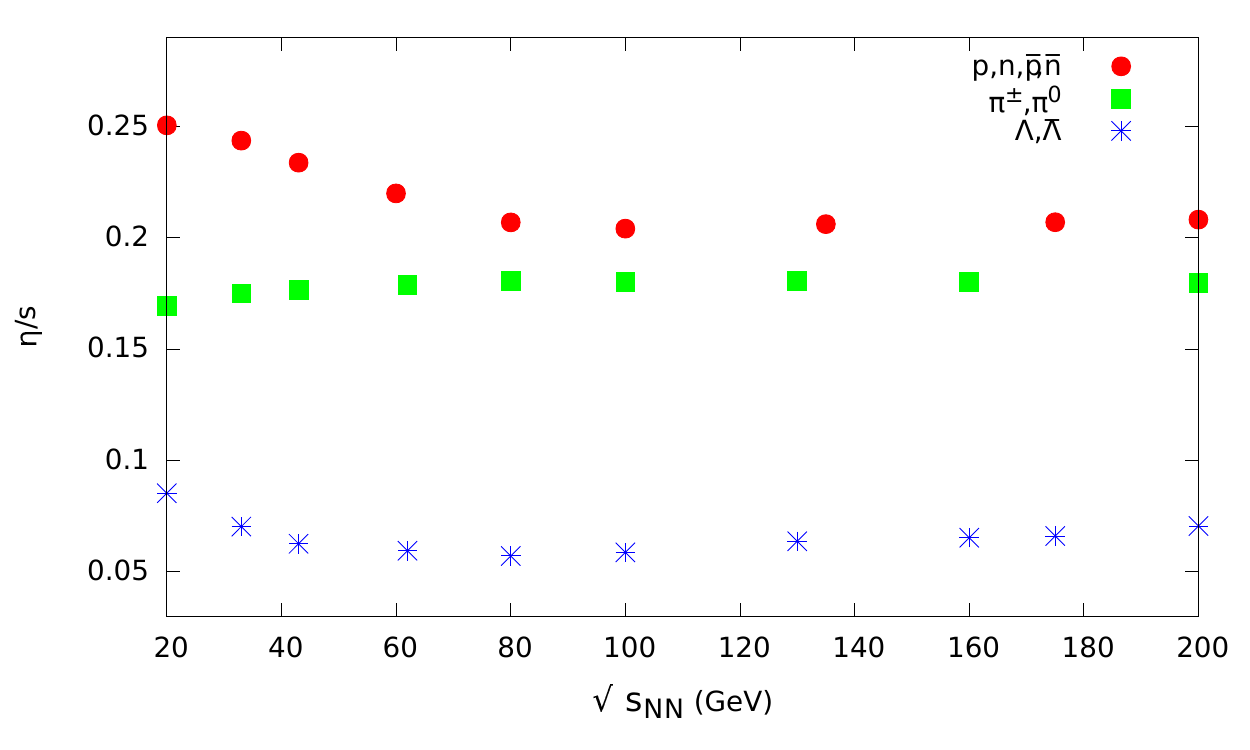}
\caption{ The specific shear viscosity at different collision energies for neutrons and protons, pions, $\Lambda$ hyperons and their antiparticles. }
\end{figure}

We know that vorticity diffuses through the fluid through the viscous stresses. So the spreading out of the vorticity patterns indicate that the bulk viscous pressure plays a greater role in the viscous diffusion of the vortices at low collision energies and high baryon chemical potential. To further understand this, we also look at the effect of lower collision energies on the elliptic flow. 

\subsection{The elliptic flow}
It is generally known from previous investigations that the elliptic flow is a good measure of the bulk and shear viscosities \cite{song}. The elliptic flow is suppressed in an non-ideal viscous fluid when compared to an ideal fluid. Previous studies show that the shear viscous contribution to the elliptic flow suppression far exceeds the bulk viscous contribution. Since we are studying the impact of collision energy on the shear and bulk viscosities, we obtain the elliptic flow from the hadronic data and compare it to the publicly available data from the STAR collaboration \cite{data}.
 
 The elliptic flow is defined by,
\begin{equation}
v_2 = \frac{\langle p_x^2 - p_y^2\rangle}{\langle p_x^2 + p_y^2 \rangle}
\end{equation} 
The elliptic flow ($v_2$) dependency on the transverse momentum ($p_T$) is well documented. Since we are interested to know whether the nature of the elliptic flow changes with change in collision energy, we plot the 
$v_2$ vs. $p_T$ for various different collision energies. In Fig 14, we show three of the plots covering the range of collision energies that we have studied. We find at lower $p_T$, the different collision energies indicate that the overall $p_T$ suppression is more for higher collision energies while no such conclusion can be reached for the higher $p_T$ range. 

\begin{figure}
\includegraphics[width = 86mm]{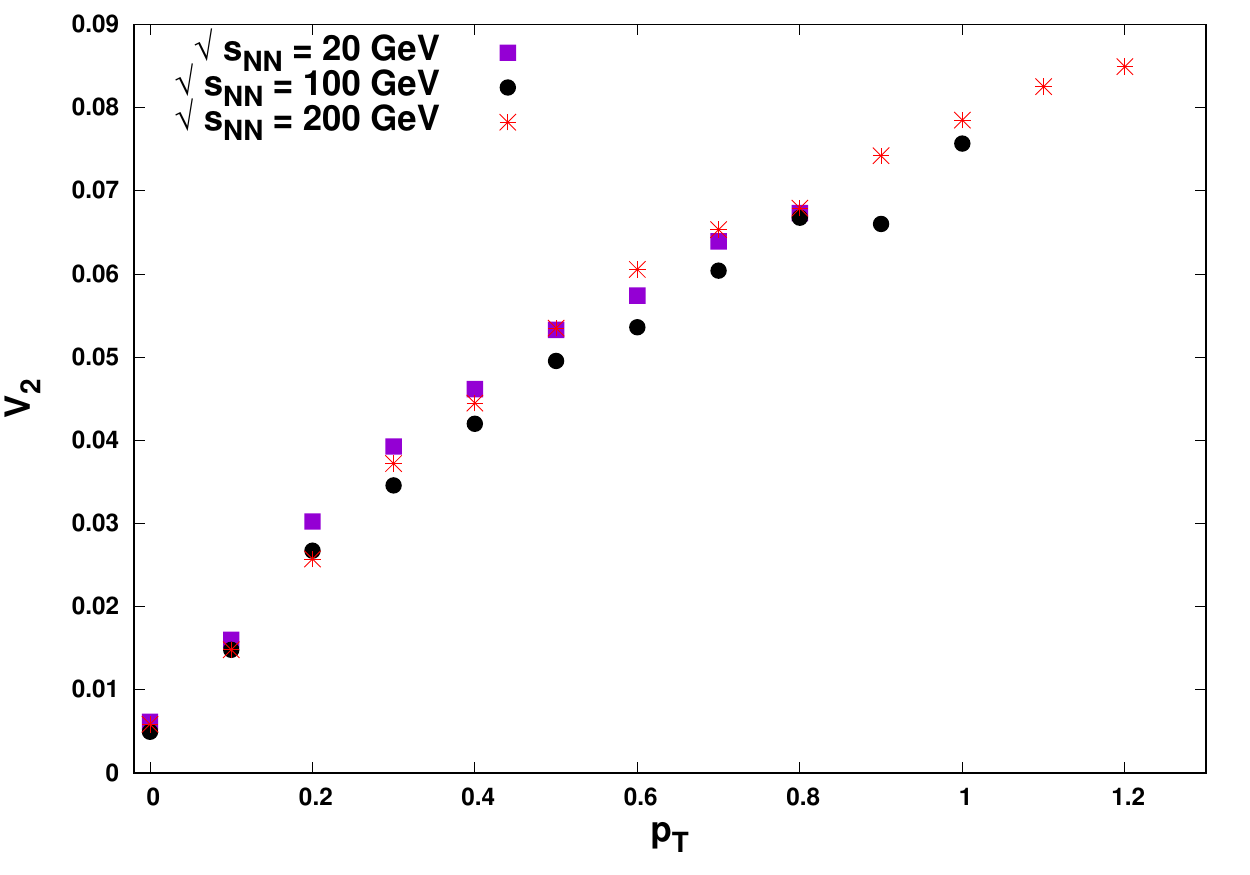}
\caption{Change of $v_2$ with $p_T$ for different collision energies}
\end{figure}

We also plot $v_2$ vs. $p_T$ for a collision energy of 19.6 GeV. The STAR collaboration has released data for  $v_2$ vs. $p_T$  for some collision energies. We plot this data along with our own calculation of  $v_2$ vs. $p_T$ for 19.6 GeV in fig 15. From the figure, it appears that the elliptic flow results from the AMPT simulations are quite close to the data obtained by the STAR collaborations at higher range of $p_T$. 

\begin{figure}
\includegraphics[width = 86mm]{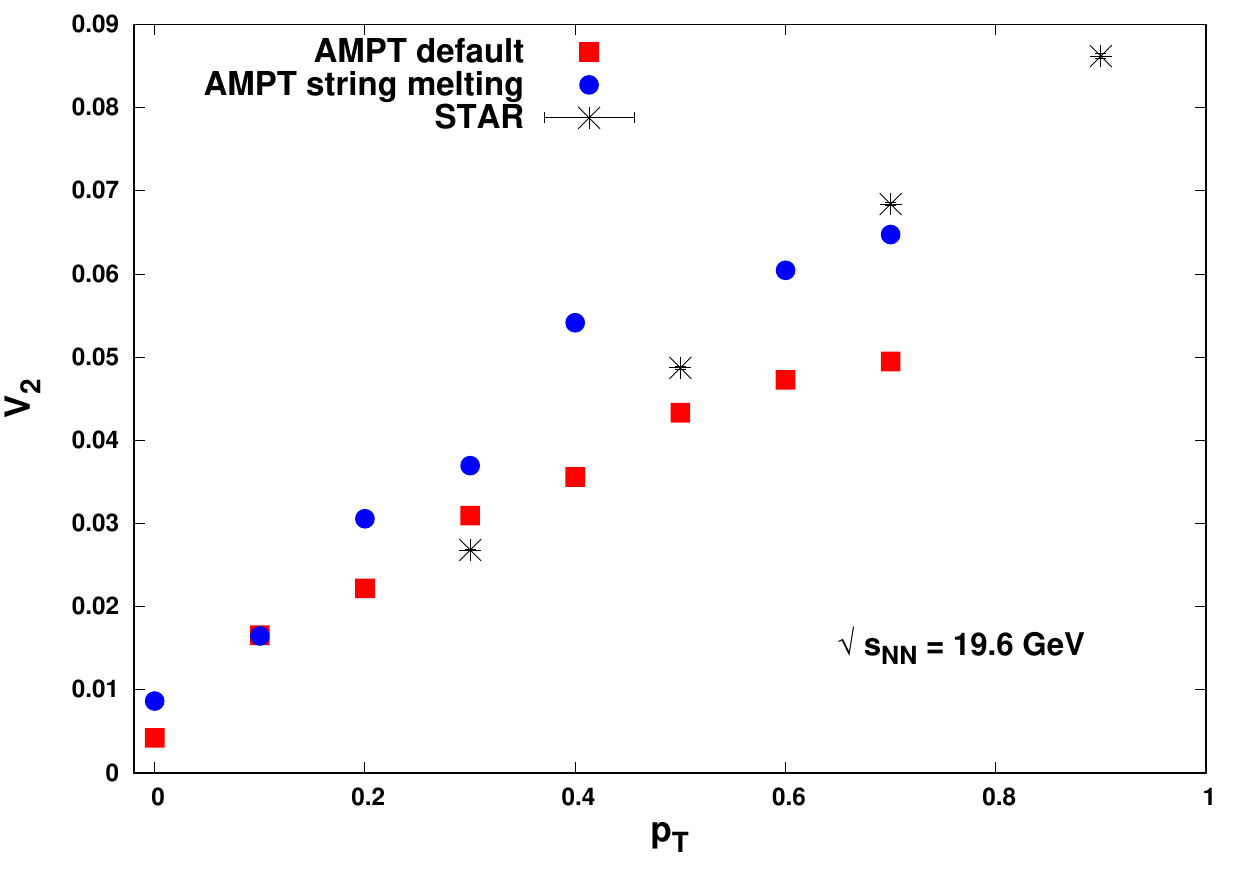}
\caption{ $v_2$ with $p_T$ at 19.6 GeV from the simulation and from data from the STAR collaboration. The data is for the 20\%-30\% centrality of charged particles. }
\end{figure}
The STAR data \cite{data} shows that the $v_2$ vs. $p_T$ plot does not change much for the collision energies between 7.7 GeV - 39 GeV. Our simulation has higher collision energies. However, even for 200 GeV we see that the $v_2$ vs. $p_T$ plot does not change significantly, when compared to the 20 GeV plot. So if shear viscosity indeed plays an important role in generating the elliptic flow, shear viscosity does not change significantly with increasing baryon chemical potential. This appears to agree with our plot of $\frac{\eta}{s}$ vs $\sqrt{s_{NN}}$ It appears that the bulk viscosity plays a greater role in the viscous effects at lower collision energies and therefore at higher chemical potentials.

\subsection{Average vorticity dependence on collision energy}
Our final results are the  dependence of the average vorticity on the collision energy. As mentioned before, we have obtained the vorticity patterns at various collision energies and presented only a few of the detailed results. We now calculate the average vorticity denoted by $\langle \omega_{xz} \rangle $ at different collision energies.  
\begin{figure}
\includegraphics[width = 86mm]{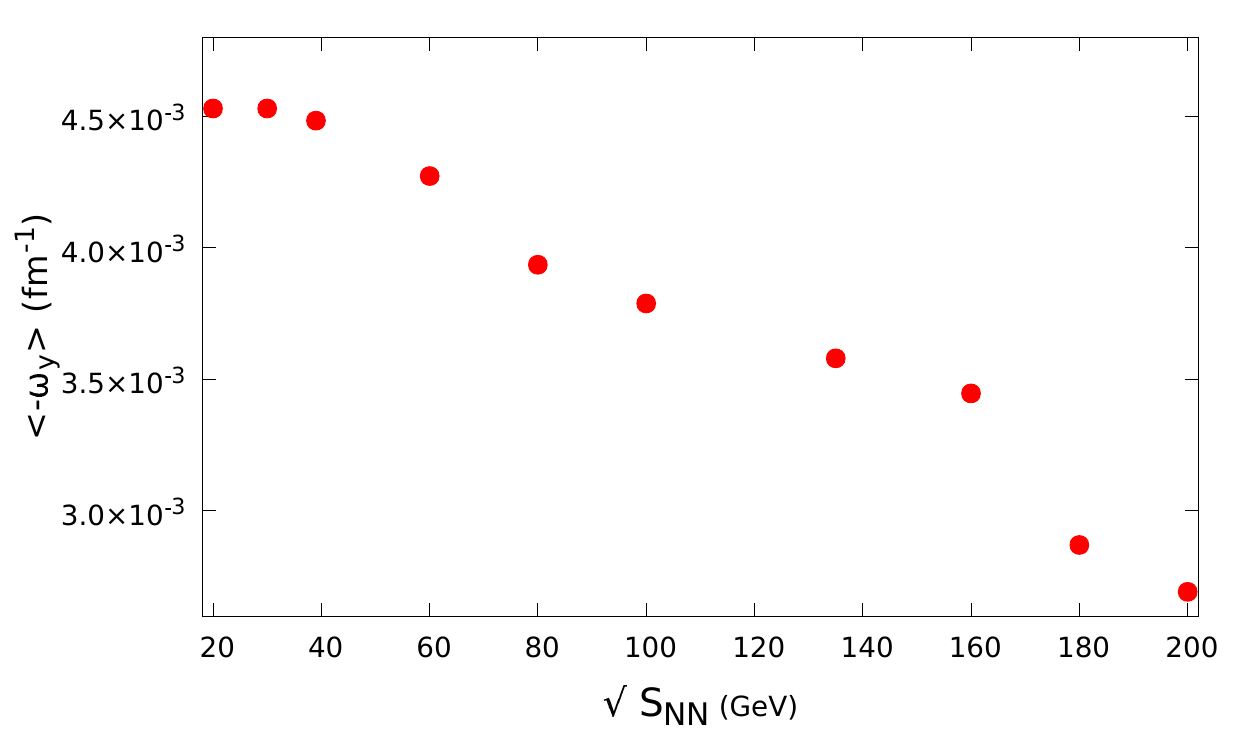}
\caption{ Average vorticity $\langle \omega_{xz} \rangle \equiv \langle \omega_{y} \rangle  $ at different collision energies ($\sqrt{s_{NN}}$) at a fixed impact parameter of $b = 7$ fm} 
\end{figure}

Our results show that the average kinetic vorticity decreases with the increase in collision energy. Our vorticity results seem to be consistent with ref.\cite{deng}. Similar plots are obtained for the relativistic and the thermal vorticities. 
In the relativistic case, we get a lot of fluctuations in the vorticity pattern. This makes the average velocity calculations quite difficult.
\begin{figure}
\includegraphics[width = 86mm]{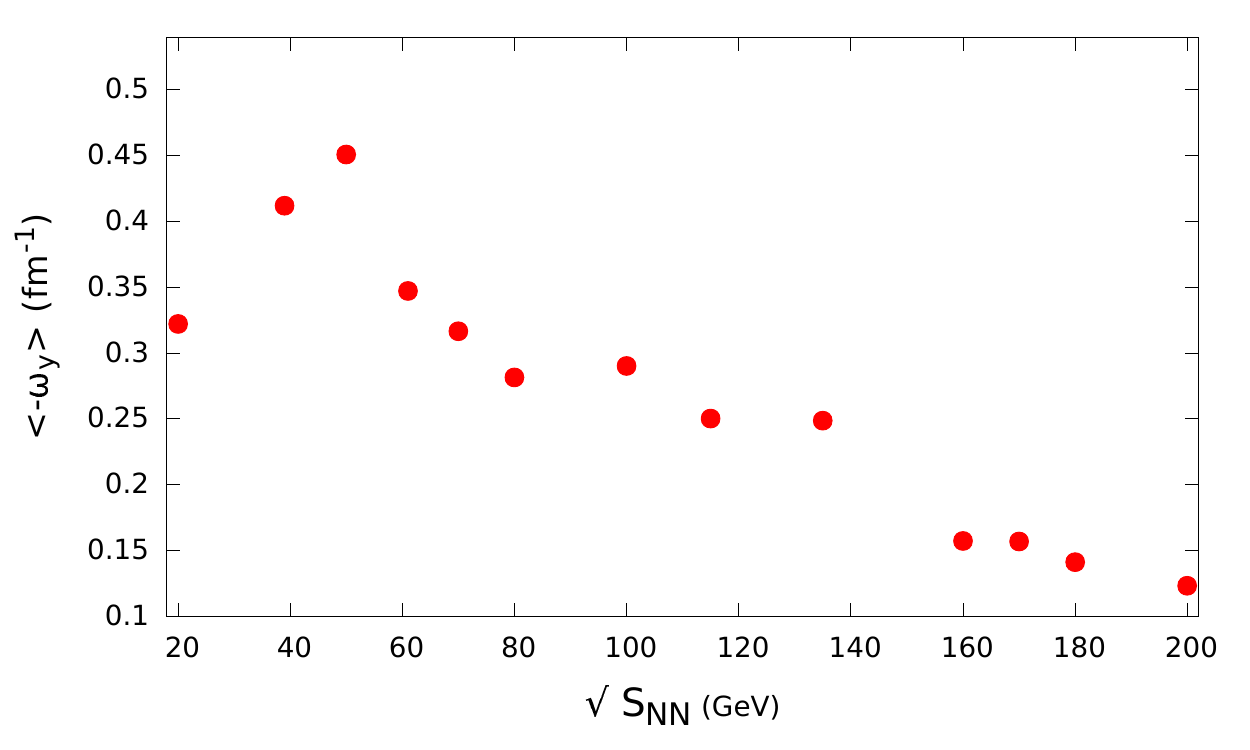}
\caption{Average vorticity $\langle \omega_{xz} \rangle \equiv \langle \omega_{y} \rangle  $  with the relativistic definition of velocities, at different collision energies ($\sqrt{s_{NN}}$) at a fixed impact parameter of $b = 7$ fm }
\end{figure}

We have however, tried to obtain the averages over the same range of collision energies. The general nature of the points remain the same as in the non relativistic case and we can thus reach the same conclusions as before. The only difference observed for the thermal and the relativistic case is that the average vorticity has a small dip below $40$ GeV. 
\begin{figure}
\includegraphics[width = 86mm]{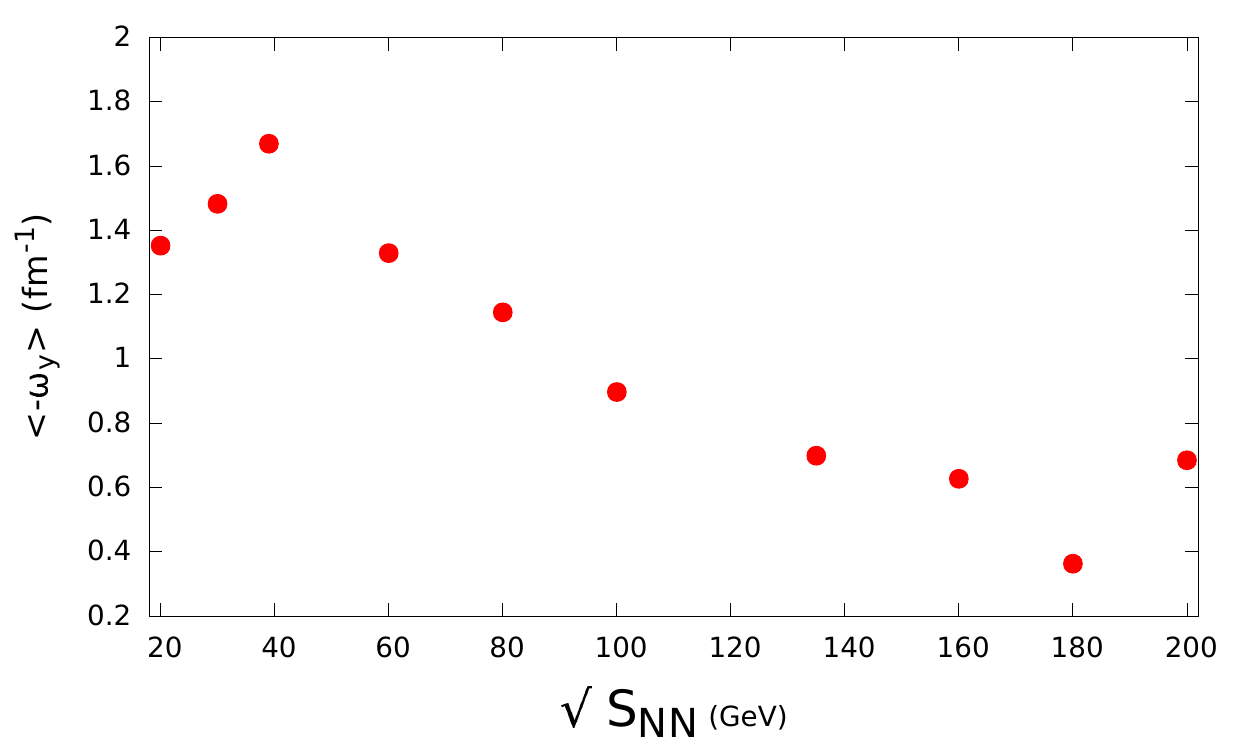}
\caption{Average thermal vorticity $\langle \omega_{xz} \rangle \equiv \langle \omega_{y} \rangle  $ at different collision energies ($\sqrt{s_{NN}}$) at a fixed impact parameter of $b = 7$ fm }
\end{figure}

As mentioned previously, the collision energies are related to the baryon chemical potential. Our plots indicate that the evolution of the vorticity pattern is better studied at lower collision energies since the average value of the vorticity is higher there for both kinetic and thermal vorticity. The bulk viscosity plays a greater role than the shear viscosity. The change in the shear viscosity is very small compared to the change in the vorticity patterns. Hence it looks like the bulk viscosity plays an important role in the evolution of the vorticity patterns in heavy ion collisions.

\section{Conclusions}

We have done a detailed study of viscous effects at different collision energies using a hybrid transport model. Such studies for a higher collision energies have been done previously. We study the vorticity patterns and viscous effects at lower collision energies also. One of the reasons why lower collision energies are studied experimentally is to include a finite baryon chemical potential. Hadron resonance gas models have been used to model the quark gluon plasma at finite baryon chemical potentials. We are interested to see whether these models can account for the vorticity patterns obtained from the hybrid transport models. Our bridge between these two very different models is the coefficient of shear viscosity.

Specifically we have looked at two definitions of vorticity. The first is the kinetic vorticity and the second is the thermal vorticity. We find that at high collision energies, the local vorticity patterns are circular and well defined. At lower collision energies they appear to be stretched and elliptical in shape. The shape of the local vorticity patterns seem to indicate that viscous stress is higher at lower collision energies where the baryon chemical potential is more. Circular vorticity patterns indicate that the angular momentum dominates over the viscous stress. At higher collision energies, chemical potential is low, we can infer that the viscous stress will also be low, our local vorticity pattern is circular showing the dominance of the angular momentum. At lower collision energy, chemical potential is high, so  the viscous stress will also be high, this gives rise to the stretched elliptical vortices as we see in the figures.

Interestingly, in the relativistic case, the local vorticity has far more fluctuations and the patterns at the higher energy collisions are not the same in the classical and the relativistic case. In the lower collision energy, however the classical and relativistic case show similar patterns but different magnitudes. We have weighted the vorticity to obtain the average vorticity at different collision energies. We find that the average vorticity goes down with increasing collision energies. This is as expected. However, our detailed patterns show that though the average vorticity turns out to be similar,  the actual vorticity patterns might have several differences. This is especially true for lower collision energy which translates to a higher baryon density. Apart from the detailed vorticity patterns we have obtained for the different definitions of vorticity, another new result is the calculation of the coefficient of shear viscosity. We have used the definition of shear viscosity from the HRG models and calculated the values at different collision energies. The coefficient of shear viscosity depends on the momentum of the particles. We find that the shear viscosity decreases very little when we change the collision energy. The same conclusion can also be reached by the study of the elliptic flow as it is related to the shear viscosity of the fluid. We find that the specific shear viscosity does not change significantly at lower collision energies. This is also corroborated by the analysis of the STAR data.

We also know that the Reynolds number is very high at these velocities. Our simulations are thus inconclusive about many aspects of viscous flow, especially whether turbulence develops or not. We plan to pursue these and various other avenues to understand the nature of the viscous nature of the quark gluon plasma that is formed in the heavy ion collisions.     

We have used the definition of shear viscosity from the hadron resonance gas model to understand the results obtained from the simulations based on the hybrid transport model. Similar use of the hadron resonance gas model has previously been done to illustrate the chemical freeze-out results obtained from the transport model study \cite{Xu}. In that case too, the transport model was the AMPT model which we have also used in our simulations. We therefore conclude that the viscous hydrodynamic results from the hadron resonance gas model and the hybrid transport models compliment each other.

\begin{center}
 Acknowledgments
\end{center} 
The authors would like to acknowledge discussions with Chitrasen Jena. For computational infrastructure, we acknowledge the Center for Modeling, Simulation and Design (CMSD) at the University of Hyderabad, where part of the simulations were carried out. A.S is supported by INSPIRE Fellowship of the Department of Science and Technology (DST) Govt. of India, through Grant no:IF170627

\end{document}